\documentclass[pra,a4paper,twocolumn,showpacs]{revtex4}

\usepackage{amssymb}
\usepackage{graphicx}
\usepackage{graphics}
\usepackage{amsmath}
\usepackage{amsbsy}
\usepackage{amsthm}
\usepackage{bbm}
\usepackage{bm}
\usepackage{epsfig}
\begin{document}

\title{Persistence Length of DNA Macromolecule with Kinks}

\author{Kyrylo~Simonov$^{1}$}
\email{kyrylo.simonov@univie.ac.at}
\author{Sergey~N.~Volkov$^{2}$}
\email{snvolkov@bitp.kiev.ua}

\affiliation{
$^1$ Fakult\"{a}t f\"{u}r Physik, Universit\"{a}t Wien, Boltzmanngasse 5, 1090 Wien, Austria\\
$^2$ Bogolyubov Institute for Theoretical Physics, 14-b Metrolohichna Str., 03680 Kyiv, Ukraine
}
\date{\today}
\pdfinfo{
  /Title    (Persistence Length of DNA Macromolecule with Kinks)
  /Author   (Kyrylo Simonov, Sergey N. Volkov)
  /Creator  ()
  /Producer ()
  /Subject  (Biomolecules)
  /Keywords (deformation, double helix, kink, Kratky-Porod model, persistence length)
}

\begin{abstract}
The study of configurational parameters of deformed DNA is a relevant problem in research of such important biological process as double helix compactization in cell. The deformations accompanied with local disruptions of the regular macromolecule structure cause significant bending of the double helix, or kinks. In this paper an approach for Kratky-Porod model to calculate persistence length of DNA macromolecule with kinks is developed. The presented approach considers kinks of arbitrary configuration, including two basic types of kinks, type 1 --- sharp kink caused by unstacking a single base pair step, and type 2 --- intrinsic-induced kink that involves several base pairs. Within developed approach analytical expressions for persistence length, coil size and gyration radius of kinky double helix were obtained.
\end{abstract}

\maketitle

\section{Introduction}

DNA double helix still remains to be one of the most attractive objects of biological physics and molecular biology since its structure was discovered by Watson and Crick. Such interest is caused by DNA function as self-replicating informational carrier that determines the structure and functioning of living organisms. Meanwhile, changes in configuration of DNA sufficiently affect its genetic activity, thus any deformations and defects in double helix structure play a key role in many biological processes, such as protein-DNA interaction, DNA packaging in chromatin and others~\cite{prevost}.

Investigations of various deformations of DNA macromolecule have been becoming very relevant lately since computational and physical experiments have demonstrated noticeable flexibility of DNA under certain biological processes. Particularly, RecA filament is able to keep DNA double helix stretched in one and a half times longer, while certain proteins can cause formation of localized extreme bends (kinks) in DNA structure~\cite{prevost}. On the other hand, DNA is sufficiently long macromolecule (for instance, human DNA has contour length of about 2 meters~\cite{alberts}), that takes form of coil in solution. So it is possible to use methods of statistical physics (e.g. idealized models of chain molecules) to describe configurations of DNA chain~\cite{grosberg} and accordingly its deformations. However, appropriate framework is developed for intact macromolecules only. In reality, long macromolecule always has some injuries that need to be determined. Therefore, defining the state of DNA double helix with some destructive factors is one of the key problems of biological physics. In particular, it is too important to take into account formation of kinks in DNA structure occuring by a wide variety of processes as intercalation of small molecules~\cite{reymer}, conformational changes~\cite{chen}, packaging~\cite{richmond} and others. 

The basic mechanical model of DNA is Kratky-Porod model (and its continous version --- WLC)~\cite{kratky} that is used as a suitable tool to well-describe wide range of mechanical properties of DNA~\cite{vologodskii}. This model is focused on a coil of smoothly curved strand, the direction of curvature at any point of strand being random~\cite{cantor}~\cite{flory}. Furthermore, it operates with several configurational parameters, that can be measured by hydrodynamical experiments and characterize state of DNA coil in solution and its flexibility. In particular, persistence length ($A$) is used as a basic parameter to characterize stiffness of DNA~\cite{grosberg}. Persistence length is a measure of distance over which DNA chain 'keeps' the direction of first segment, so correlation between directions of two segments decreases exponentially (with a typical length $A$) while increasing a contour length between them~\cite{cantor}~\cite{bustamante}. Particularly, DNA double helix has a persistence length of $A \approx 500$\r{A}.

Meanwhile, Kratky-Porod model uses harmonic approach so the energy cost of bending ($E$) depends quadratically on bending angle ($\tilde{\delta}$)~\cite{vologodskii}. In that way bending rigidity is simply proportional to persistence length: $k = k_B T A$, where $k_B$ is a Boltzmann constant and $T$ is a temperature~\cite{bustamante}. Therefore, it is expected that Kratky-Porod model describes smooth deformations of double helix with relatively small changes between the bonds. So arises the problem of describing DNA with kinks of various nature in the framework of Kratky-Porod model. This problem has become very important over the past decade due to the numerous computational experiments with various types of kinks, e.g. studies of DNA cyclization~\cite{lankas}. Furthermore, Wiggins and colleagues proposed an extension of WLC model -- KWLC, that takes into account sharp kinks characterized by probability of kinking per unit length~\cite{wiggins}. However, kinks of KWLC model are only of one type and, generally speaking, idealized. Therefore, in this paper we present a development of Kratky-Porod model to describe formation of kinks in DNA structure with different length and configuration and propose a way to calculate several configurational parameters that describe flexibility of DNA and state of its coil, specifically persistence length ($A$), coil size ($R$) and gyration radius ($G$).

\section{Presenting the kinks}

\begin{figure}[ht!]\centering\includegraphics[width=5cm]{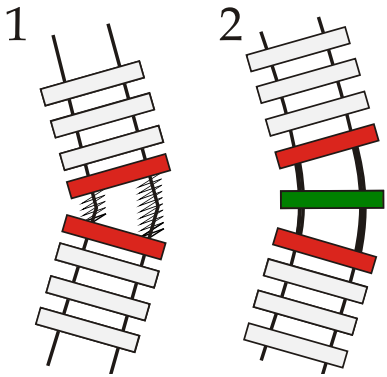}\caption{(Colour online) Diagrammatic representation of two types of kinks in double helix: 1) type 1 --- single-stranded break, 2) type 2 --- intrinsic-induced kink. Red colour corresponds to intact base pairs, green one represents modified base pair.}\label{KinkTypes}\end{figure}

Computer simulations of DNA minicircles detected two basic groups of kinks depending on their nature~\cite{lankas}. Type 1 presents sharp kink caused by unstacking a single base pair step. Such kink looks like proposed by Crick and Klug extreme local deformation with sharp loss of stacking~\cite{crick} and involves a strong bend between two intact consecutive base pairs~\cite{lankas} with a local disruption of link, as shown on Fig.~\ref{KinkTypes}. Generally speaking, kinks of type 1 are rather idealized, however similar damages of DNA structure can be caused by intercalation of small molecules into double helix~\cite{reymer}~\cite{chen}~\cite{saenger}.

On the other hand, type 2 presents intrinsic-induced kink with changed stacking and distributed over two base-pair steps with modified central base pair~\cite{prevost}~\cite{lankas}, as shown on Fig.~\ref{KinkTypes}. Such kink is suggested as more probable~\cite{prevost} than kink of type 1, and several processes as changing of hydrogen bonding~\cite{volkov1995}~\cite{kryachko}, flipping into Hoogsteen base pairs~\cite{saenger} and others are good candidates to cause such kinks.

Strictly speaking, other types of kinks may exist. In particular, conformational transformation (e.~g.~B-A transformation in double helix~\cite{volkov}), binding of TBP to TATA-box~\cite{klug}~\cite{kanevska} could cause conformational kink with similar to type 2 structure, but involving more than three base pairs. Therefore we can besides consider conformational kinks distributed on more than three base pairs, saying, four base pairs.

\begin{figure}[ht!]\centering\includegraphics[width=5cm]{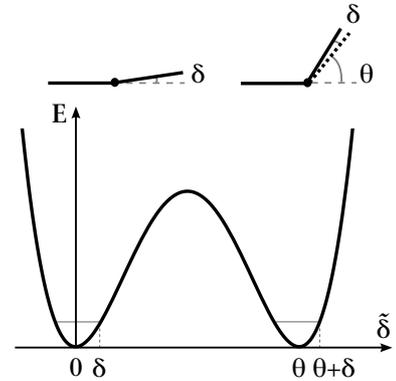}\caption{Top: two possible states of monomeric element, normal (with bending angle $\delta$) and kinky (with bending angle $\vartheta + \delta$). Bottom: minimums of double well represent two states of monomeric element.}\label{DoublePotential}\end{figure}

Each kink can be characterized by its total angle ($\vartheta$) and phenomenological probability of its formation ($W$) that represents in fact concetration of kinks in double helix and, generally speaking, depends on several parameters, e.g. bending energy $E$, spring constant $k$ (for example, KWLC model presents kinking probability as $W = 2ke^{-E}$ for kink of type 1~\cite{wiggins}) etc. As bending energy in Kratky-Porod model is calculated in harmonic approach by small bending angle ($\delta$), we propose to use double well potential~\cite{krumhansl} that represents two possible states of monomeric element: normal and kinky, as shown on Fig.~\ref{DoublePotential}.
Monomeric element in normal state is bent at small angle ($\delta$) caused by thermal fluctuations just as in Kratky-Porod model. On the other hand, kinky state of monomeric element represents formation of kink with certain angle ($\vartheta$). Therefore we can represent total bending angle ($\tilde{\delta}$) of monomeric element in kinky state as the sum of kink angle ($\vartheta$) and small deviation ($\delta$):
\begin{eqnarray}
\nonumber
& \tilde{\delta} = \vartheta \pm \delta,\\
& \cos \tilde{\delta} = \cos \vartheta \cos \delta \mp \sin \vartheta \sin \delta \approx \cos \vartheta \cos \delta.
\end{eqnarray}
With use of such representation of two possible states we can accordingly modify persistence length and other configurational parameters of DNA to consider the effect of kink formation and its influence on the state of DNA coil.

\section{Kinky persistence length}

\begin{figure}[ht!]\centering\includegraphics[width=7cm]{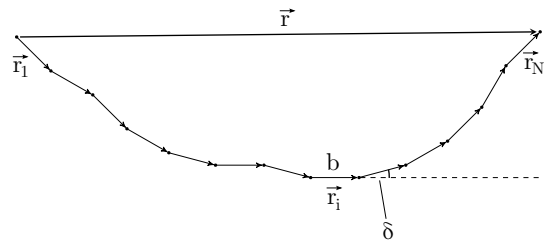}\caption{Diagrammatic representation of Kratky-Porod chain.}\label{KPChain}\end{figure}Kratky-Porod model represents persistence length of macromolecular chain as a limit of average sum of projections of segment vectors ($\vec{r}_i$) onto direction of the first segment ($\vec{r}_1$)~\cite{cantor}~\cite{hagerman}:
\begin{eqnarray}
A = \lim_{N \rightarrow \infty} \langle \frac{\vec{r}_1}{b} \cdot \vec{r} \rangle = b \lim_{N \rightarrow \infty} \sum_{i=0}^{N-1} \langle \cos \delta \rangle ^i = \frac{b}{1 - \langle \cos \delta \rangle},
\end{eqnarray}
where $b$ is a length of each segment, $N$ is a number of segments, $\vec{r} = \sum_{i=1}^{N} \vec{r}_i$ is end-to-end vector, as shown on Fig.~\ref{KPChain}. Such definition of persistence length represents a chain with segments being all in normal state.

\subsection{Kinks of type 1}

We start to construct our approach with kinks of type 1 that have the simplest structure. According to definition of persistence length we review all the series of chain segments and introduce for each one a probability of kink formation ($W$):
\begin{eqnarray}
\langle \cos \delta \rangle \rightarrow \bigl(1 - W \bigl(1 - \cos\vartheta \bigr)\bigr) \langle \cos \delta \rangle.
\end{eqnarray}

Now, collecting modified contributions of all segments, we can calculate persistence length of chain with kinks of type 1:
\begin{eqnarray}
\mathcal{A}_1 = \frac{b}{1- \bigl(1 - W \bigl( 1- \cos\vartheta \bigr) \bigr) \langle \cos \delta \rangle}.
\end{eqnarray}

\subsection{Kinks of type 2 and conformational kinks}
Kink of type 2 is distributed over two segments of chain, so we introduce probability of such kink formation in more complicated way:
\begin{eqnarray}
\displaystyle
& \langle \cos \delta \rangle + \langle \cos \delta \rangle ^2 \rightarrow
\\
\nonumber & \bigl(1-W\bigl(1-\cos \frac{\vartheta}{2} \bigr)\bigr) \langle \cos \delta \rangle +
\\
\nonumber & \bigl(1-W\bigl(1-\cos ^2 \frac{\vartheta}{2} \bigr)\bigr) \langle \cos \delta \rangle ^2.
\end{eqnarray}
Summing all the contributions we obtain expression for persistence length of chain with kinks of type 2:
\begin{eqnarray}
\mathcal{A}_{2} = \frac{b \chi_2 \bigl(W, \vartheta \bigr)}{1- \bigl(1 - W \bigl( 1- \cos ^2 \frac{\vartheta}{2} \bigr) \bigr) \langle \cos \delta \rangle ^2},
\end{eqnarray}
where $\chi_2 \bigl(W, \vartheta \bigr) = 1 + \bigl(1-W\bigl(1-\cos \frac{\vartheta}{2} \bigr)\bigr) \langle \cos \delta \rangle$ characterizes segments involved by kink of type 2.

On the other hand, we can focus on conformational kinks with resembling structure as kinks of type 2, but involving three base pairs:
\begin{eqnarray}
\displaystyle
& \langle \cos \delta \rangle + \langle \cos \delta \rangle ^2 + \langle \cos \delta \rangle ^3 \rightarrow
\\
\nonumber & \bigl(1-W\bigl(1-\cos \frac{\vartheta}{3} \bigr)\bigr) \langle \cos \delta \rangle +
\\
\nonumber & \bigl(1-W\bigl(1-\cos ^2 \frac{\vartheta}{3} \bigr)\bigr) \langle \cos \delta \rangle ^2 +
\\
\nonumber & \bigl(1-W\bigl(1-\cos ^3 \frac{\vartheta}{3} \bigr)\bigr) \langle \cos \delta \rangle ^3.
\end{eqnarray}
In that way persistence length of chain with conformational kinks can be expressed as:
\begin{eqnarray}
\mathcal{A}_{3} = \frac{b \chi_3 \bigl(W, \vartheta \bigr)}{1- \bigl(1 - W \bigl( 1- \cos ^3 \frac{\vartheta}{3} \bigr) \bigr) \langle \cos \delta \rangle ^3},
\end{eqnarray}
where $\chi_3 \bigl(W, \vartheta \bigr) = 1 + \bigl(1-W\bigl(1-\cos \frac{\vartheta}{3} \bigr)\bigr) \langle \cos \delta \rangle + \bigl(1-W\bigl(1-\cos^2 \frac{\vartheta}{3} \bigr)\bigr) \langle \cos \delta \rangle^2$ characterizes segments involved by conformational kink.

\subsection{Kinks of arbitrary form}

At last, we can generalize obtained approach by focusing on hypothetic kinks with arbitrary length and distribution of angles inside the kink. Accordingly persistence length of chain with such kinks can be defined as:
\begin{eqnarray}
\mathcal{A}_{d} = \frac{b \chi_d \bigl(W, \vartheta_k \bigr)}{1- \bigl(1 - W \bigl( 1- \prod\limits_{k=1}^d \cos \vartheta_k \bigr) \bigr) \langle \cos \delta \rangle ^d},
\end{eqnarray}
where $\chi_d \bigl(W, \vartheta_k \bigr) = 1 + \sum\limits_{i=1}^{d-1} \bigl(1 - W \bigl( 1- \prod\limits_{k=1}^i \cos \vartheta_k \bigr) \bigr) \langle \cos \delta \rangle ^i$ characterizes segments involved by such hypothetic kink.

\section{Kinky coil size and gyration radius}

Coil size and gyration radius can be modified to take formation of kinks into account in much simpler way than persistence length, so it is necessary to replace persistence length with kinky one in corresponding expressions only. Thus kinky coil size is defined as:
\begin{eqnarray}
\langle \mathcal{R}_{d} \rangle^2 = [\langle R \rangle^2]_{A\rightarrow \mathcal{A}_d} = 2 \mathcal{A}_d^2 \Bigl(\frac{L}{\mathcal{A}_d} - 1 + e^{-L/\mathcal{A}_d}\Bigr),
\end{eqnarray}
where $L$ is a contour length of chain. In the same way gyration radius is defined as:
\begin{eqnarray}
& \langle \mathcal{G}_{d} \rangle^2 = [\langle G \rangle^2]_{A\rightarrow \mathcal{A}_d} = 
\\
\nonumber & \frac{L\mathcal{A}_d}{3} - \mathcal{A}_d^2 + \frac{2\mathcal{A}_d^3}{L} - \frac{2\mathcal{A}_d^4}{L^2} (1 - e^{-L/\mathcal{A}_d}).
\end{eqnarray}

\section{Discussion}
\begin{figure}[ht!]\centering\includegraphics[width=8.5cm]{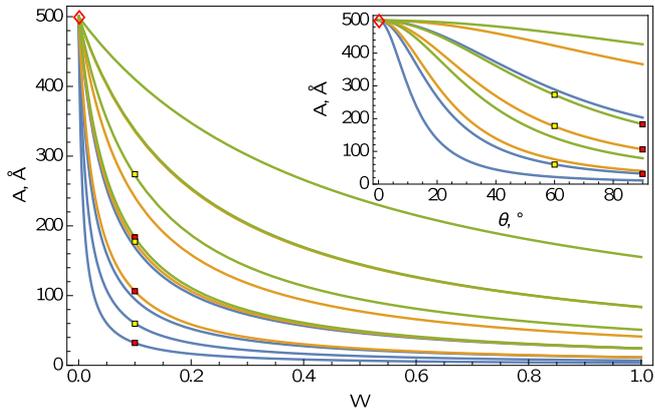}\caption{(Colour online) Sharp decreasing of persistence length $A$ is caused by increasing the probability of kink formation $W$. Different curves correspond to fixed angle $\vartheta = 30^{\circ}, 45^{\circ}, 60^{\circ}, 90^{\circ}$ (from top to down). Inset: persistence length $A$ vs angle of kink $\vartheta$ with fixed formation probability $W = 0.01, 0.1, 0.3$ (from top to down). Blue line represents kinks of type 1, orange one represents kinks of type 2, green one represents conformational kinks and red diamond corresponds to normal persistence length $A=500$\r{A} on both plots. Red and yellow rectangles represent results for $W=0.1$, $\vartheta = 90^{\circ}$ and $W=0.1$, $\vartheta = 60^{\circ}$ accordingly.}\label{Results_PersLength}\end{figure}

We now have an approach for Kratky-Porod model that describes persistence length of chains with kinks characterized by two parameters, kink angle ($\vartheta$) and kink formation probability ($W$). Let us begin a discussion of results by calculating persistence length of chains with three types of kinks: type 1, type 2 and conformational kinks. Since we investigate dsDNA, we are assuming that $A=500$\r{A} and $b=3.4$\r{A}.

While Fig.~\ref{Results_PersLength} represents changing of persistence length under increasing probability ($W$) or angle ($\vartheta$), it is clear that stiffness of DNA macromolecule decreases very sharply with changing of kink parameters in direction of bigger values. For example, taking into account formation of kinks with $W=0.1, \vartheta=90^{\circ}$ we find that persistence length decreases by $94\%$ for type 1, $79\%$ for type 2 and $63\%$ for conformational kinks. Thus results shown on Fig.~\ref{Results_PersLength} strongly suggest that even moderate possibility of kinks formation in DNA structure directly changes conformational properties of macromolecule and its flexibility. On the other hand, with increasing the size of kink (concerning conformational kink, for example) stiffness of DNA approaches to initial value again. In other words, macromolecule 'feels' conformational kinks much weaker than kinks of type 1: decreasing of persistence length by presence of conformational kinks with $W=0.1, \vartheta=90^{\circ}$ is $32\%$ down on decreasing by kinks of type 1 with the same parameters. Therefore changing of kink configuration can decrease DNA flexibility as well as increase.

Summarizing, we have presented in this paper a simple framework for Kratky-Porod model that allows to calculate directly persistence length of DNA macromolecule with kinks of certain type occured in its structure. Despite its simplicity our approach being characterized by total kink angle ($\vartheta$) and probability of kink formation ($W$) describes flexibility of DNA chain with kinks of different possible intrinsic structure and length $d$, contrasting to KWLC model that describes kinks of type 1 only~\cite{wiggins}. In particular, it is important to focus on two main types of kinks being investigated in computational experiments~\cite{prevost}~\cite{lankas} since corresponding configurational parameters of DNA can be measured and compared with predictions of the approach. Moreover, it is possible to define concenration of kinks in DNA chain and their nature by analyzing predicted configurational parameters.

Since our approach is quite simple, it would be desirable in future perspective to extend it to more complicated configurations of DNA macromolecule, e.~g. with kinks of few types in its structure. We expect that such generalization of proposed approach would be applicable to various kinky macromolecules as a realistic theoretical tool to analyze their flexibility and conformational properties. Thus the ability of our approach to describe deformations of various nature and structure is a good starting point to build a more complicated model of kinky DNA macromolecule involved to key biological processes such as folding, transcription etc.


\begin{thebibliography}{1}

\bibitem{prevost}
C.~Pr\'{e}vost, M.~Takahashi and R.~Lavery, ChemPhysChem \textbf{10}, 1399 (2009)
\bibitem{alberts}
B.~Alberts, A.~Johnson, J.~Lewis et al., ``{\em Molecular Biology of Cell}'', (4th Edition, Garland Science) (2002)
\bibitem{grosberg}
A.~Yu.~Grosberg and A.~R.~Khokhlov, ``{\em Statistical Physics of Macromolecules}'', (AIP Press) (1994)
\bibitem{reymer}
A.~Reymer and B.~Nord\'{e}n, Chem. Commun. \textbf{48}, 4941 (2012)
\bibitem{chen}
C.-Y.~Chen, T.-P.~Ko, T.-W.~Lin et al., Nucl. Acids Res. \textbf{33}, 430 (2005)
\bibitem{richmond}
T.~J.~Richmond and C.~A.~Davey, Nature \textbf{423}, 145 (2003)
\bibitem{kratky}
O.~Kratky and G.~Porod, Recl. Trav. Chim. Pays-Bas \textbf{68}, 1106 (1949)
\bibitem{vologodskii}
A.~Vologodskii and M.~D.~Frank-Kamenetskii,  Nucl. Acids Res. \textbf{41}, 6785 (2013)
\bibitem{cantor}
C.~R.~Cantor and P.~R.~Schimmel, ``{\em Biophysical Chemistry,
Part III: The Behavior of Biological Macromolecules}'', (W. H. Freeman) (1980)
\bibitem{flory}
P.~J.~Flory, ``{\em Statistical Mechanics of Chain Molecules}'', (Interscience Publishers) (1969)
\bibitem{bustamante}
C.~Bustamante, Z.~Bryant and S.~B.~Smith, Nature \textbf{421}, 423 (2003)
\bibitem{lankas}
F.~Lanka\v{s}, R.~Lavery and J.~H.~Maddocks, Structure \textbf{14}, 1527 (2006)
\bibitem{wiggins}
P.~A.~Wiggins, R.~Phillips and P.~C.~Nelson, Phys. Rev. E \textbf{71}, 021909 (2005)
\bibitem{crick}
F.~H.~C.~Crick and A.~Klug, Nature \textbf{255}, 530 (1975)
\bibitem{volkov1995}
S.~N.~Volkov, Molek. Biol. (Moscow) \textbf{29}, 1086 (1995)
\bibitem{kryachko}
E.~S.~Kryachko and S.~N.~Volkov, Int. J. Quantum Chem. \textbf{82}, 193 (2001)
\bibitem{saenger}
W.~Saenger, ``{\em Principles of Nucleic Acid Structure}'', (Springer-Verlag) (1984)
\bibitem{volkov}
S.~N.~Volkov, J. Biol. Phys. \textbf{31}, 323 (2005)
\bibitem{klug}
A.~Klug, Nature \textbf{365}, 486 (1993)
\bibitem{kanevska}
P.~P.~Kanevska and S.~N.~Volkov, Ukr. J. Phys. \textbf{51}, 1003 (2006)
\bibitem{krumhansl}
J.~A.~Krumhansl and J.~R.~Schrieffer, Phys. Rev. B \textbf{11}, 3535 (1975)
\bibitem{hagerman}
P.~J.~Hagerman, Ann. Rev. Biophys. Biophys. Chem. \textbf{17}, 265 (1988)
\end{thebibliography}
\end{document}